\documentclass[conference]{IEEEtran}
\usepackage{cite}
\usepackage{amsmath,amssymb,amsfonts}
\usepackage{algorithmic}
\usepackage{graphicx}
\usepackage{textcomp}
\usepackage{xcolor}
\usepackage{amsfonts}
\usepackage{multirow}
\usepackage{nicefrac}
\usepackage{tabularx}
\usepackage{enumerate}
\usepackage{tikz}
\usepackage{microtype}

\newcommand{\textapprox}{\raisebox{0.5ex}{\texttildelow}}

\usepackage[binary-units]{siunitx}
\DeclareSIUnit \dBm {dBm}
\DeclareSIUnit \dB {dB}
\DeclareSIUnit \dBi {dBi}
\DeclareSIUnit \Kbps {Kbps}
\DeclareSIUnit \Mbps {Mbps}
\DeclareSIUnit \Gbps {Gbps}
\DeclareSIUnit \kBps {kBps}
\DeclareSIUnit \MBps {MBps}
\DeclareSIUnit \GBps {GBps}

\usepackage[hyphens]{url}
\usepackage[hidelinks]{hyperref}
\hypersetup{breaklinks=true}

\usetikzlibrary{arrows,automata}
\def\BibTeX{{\rm B\kern-.05em{\sc i\kern-.025em b}\kern-.08em
    T\kern-.1667em\lower.7ex\hbox{E}\kern-.125emX}}
\begin{document}

\title{Reliable IoT Firmware Updates: A Large-scale Mesh Network Performance Investigation}

\author{\IEEEauthorblockN{Ioannis Mavromatis\IEEEauthorrefmark{1}, Aleksandar Stanoev\IEEEauthorrefmark{1}, Anthony J. Portelli\IEEEauthorrefmark{1}, Charles Lockie\IEEEauthorrefmark{2}, \\Marius Ammann\IEEEauthorrefmark{2}, Yichao Jin\IEEEauthorrefmark{1}, and Mahesh Sooriyabandara\IEEEauthorrefmark{1}}
\IEEEauthorblockA{\IEEEauthorrefmark{1} Bristol Research and Innovation Laboratory (BRIL), Toshiba Europe Ltd., Bristol, UK\\
\IEEEauthorrefmark{2} Department of Electrical and Electronic Engineering, University of Bristol, Bristol, UK\\
Emails: \{Ioannis.Mavromatis, Aleksandar.Stanoev, Yichao.Jin\}@toshiba-bril.com}}

\maketitle

\begin{abstract}
Internet of Things (IoT) networks require regular firmware updates to ensure enhanced security and stability. As we move towards methodologies of codifying security and policy decisions and exchanging them over IoT large-scale deployments (security-as-a-code), these demands should be considered a routine operation. However, rolling out firmware updates to large-scale networks presents a crucial challenge for constrained wireless environments with large numbers of IoT devices. This paper initially investigates how the current state-of-the-art protocols operate in such adverse conditions by measuring various Quality-of-Service (QoS) Key Performance Indicators (KPIs) of the shared wireless medium. We later discuss how Concurrent Transmissions (CT) can extend the scalability of IoT protocols and ensure reliable firmware roll-outs over large geographical areas. Measuring KPIs such as the mesh join time, the throughput, and the number of nodes forming a network, we provide great insight into how an IoT environment will behave under a large-scale firmware roll-out. Finally, we conducted our performance investigation over the UMBRELLA platform, a real-world IoT testbed deployed in Bristol, UK. This ensures our findings represent a realistic IoT scenario and meet the strict QoS requirements of today's IoT applications.
\end{abstract}


\section{Introduction}\label{sec:intro}
Internet of Things (IoT) has become synonymous with everyday computing. It provides the ability for interconnected devices to exchange information over a wireless medium without human intervention~\cite{IoTSurvey}. IoT systems and platforms are already utilised in home environments for voice assistants and security systems to energy, heating, and lighting control~\cite{smartHomes}. Moving forward, IoT is expected to impact commercial applications in a number of areas, including Industry 4.0, connected vehicles and smart grids~\cite{IoTbasedSmartCities}.

Operational stability and security are two critical factors of every IoT network and application. Both can be achieved with frequent firmware roll-outs~\cite{firmware_rollout}. However, autonomous firmware update without physical proximity is susceptible to adverse wireless medium conditions, can drastically increase the roll-out times, or lead to downtimes. Interconnected IoT devices make it possible for billions of nodes to exchange information and transform raw data into meaningful inferences. This was made possible due to the standardisation of IoT communication protocols. These constrained devices, having limited resources, can connect and form huge distributions of networks~\cite{Cano2019EvolutionPerspective}. Such networks can ensure massive firmware roll-outs over large geographical areas. However, the current IoT protocols were initially designed to work within home environments~\cite{Cano2019EvolutionPerspective} and do not scale well in city-wide deployments~\cite{scalabilityMAC}.

The contribution of this paper is two-fold. Initially, we investigate the requirements for such a firmware roll-out. Based on that, we explore the coexistence, cooperation, and interoperability of IoT communication protocols in a large-scale city-wide deployment. We intend on showing how different protocols perform within a real-world environment and their limitations as the number of devices increases. We also discuss how Concurrent Transmissions (CT)~\cite{baddeley2019Atomic} can extend the scalability of IoT protocols. Well-established IEEE 802.15.4 protocols such as the Carrier Sense Multiple Access with Collision Avoidance (CSMA-CA)~\cite{csma} have been widely tested and implemented in the IoT world. For our investigation, we use CSMA-CA as a benchmark. As a second step, we conduct a large-scale performance investigation using a CT-based implementation~\cite{baddeley2019Atomic}. Our analysis shows that CT can increase the current system performance in a real-world environment, especially when the number of interconnected devices increases.

Our performance investigation was based on Urban Multi Wireless Broadband and IoT Testing for Local Authority and Industrial Applications (UMBRELLA). UMBRELLA is currently deployed in Bristol, UK, and is a large-scale urban testbed. Owing to its size and versatility, it can be used as a playground for wireless experimentation and assessing Key Performance Indicators (KPIs) for large-scale urban IoT scenarios. Our investigation focuses on various KPIs like mesh creation times, reliability and data throughput, linking them with an example application of a large-scale firmware roll-out.

The rest of the paper is organised as follows. Sec.~\ref{sec:system} describes the system architecture, our scenario, and the experimental setup. Sec.~\ref{sec:trials} presents the workflow and provides a better insight on the implementations and the large-scale evaluation. Our comparative results can be found in Sec.~\ref{sec:performance}. Finally, the paper summarises our investigation in Sec.~\ref{sec:conclusion}.

\section{System Description}\label{sec:system}
Our experimentation is based on a realistic IoT scenario. We consider a use-case where a firmware update is rolled out to a wide-reaching network of IoT devices. This can be the case when, for example, a security vulnerability is found, subsequently, a patch is introduced, and the new firmware is reflashed on all the wireless IoT interfaces. Within such a scenario, the downtime on the network should be minimised while ensuring the new firmware reliably reaches all the destination nodes without significant delays. To emulate that, a \SI{100}{\kilo\byte} file is used to represent the firmware binary. This is roughly the size of an uncompressed binary file for the Nordic Semiconductor nRF52840~\cite{nRF52840} System on Chip (SoC). Data compression could be used to reduce the size of the payload substantially. However, this would require the receiving nodes to perform the decompression. Such functionality is somewhat onerous when considering the limited computing capacity of IoT devices, particularly if they are battery-powered. Therefore, the uncompressed file size was chosen for our experimentation.

\begin{figure}[t]
    \centering
    \includegraphics[width=0.98\columnwidth]{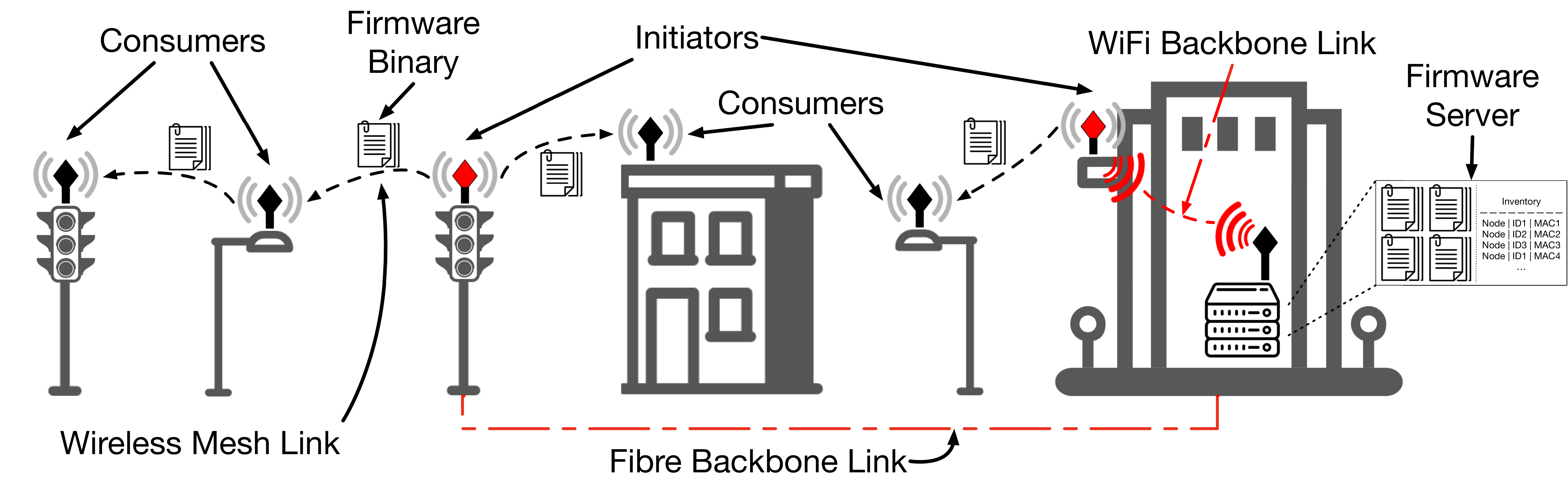}
    \vspace{-4mm}
    \caption{A high-level system architecture and its implementation on UMBRELLA. The nodes share the firmware binaries over the wireless mesh network. The initiators are reachable from the server via a backbone link.}
    \label{fig:umbrellaarchitecture}
\end{figure}

We assume that a firmware server stores the firmware images and manifests, and distributes them to the IoT devices (as in Fig.~\ref{fig:umbrellaarchitecture}). The firmware binary is sent to all firmware consumers over a wireless mesh link. What is more, the server has direct access to the roll-out initiators (source nodes) and initiates the deployment. To do so, a list of all active consumers is always stored on the server-side and can be periodically updated and maintained. In the following sections, we describe the wireless protocol stacks used for our investigation.

\subsection{IEEE 802.15.4, Bluetooth and the Different PHYs}\label{subsec:phys}
Bluetooth Low Energy (BLE) and Low Rate Wireless Personal Area Networks (LR-WPANs) (IEEE 802.15.4) have emerged as wireless communications technologies of choice in many IoT applications. We utilise both during our investigation. The most recent version (BLE 5) supports four PHYs that largely differ in terms of data rate and robustness~\cite{ble5PHY}, i.e. \SI{2}{\Mbps}, which doubles the nominal throughput of the original \SI{1}{\Mbps} PHY, and two coded PHYs with coding rates of $1/2$ and $1/8$ (i.e., the \SI{500}{\Kbps}  and \SI{125}{\Kbps} respectively). The IEEE 802.15.4 supports a datarate of \SI{250}{\Kbps} in the \SI{2.4}{\giga\hertz} frequency band and up to \SI{40}{\Kbps} at the sub-\SI{}{\giga\hertz} one. For our experimentation, we will use all four PHYs from BLE and the \SI{250}{\Kbps} PHY for IEEE 802.15.4.

Both IEEE 802.15.4 and BLE operate in the global unlicensed ISM band of \SI{2.4}{\giga\hertz}. The same frequency band is also used by wireless technologies such as IEEE 802.11. Furthermore, IEEE 802.15.4g operates in the sub-\SI{}{\giga\hertz} frequency band as well (\SIrange{868}{868.6}{\mega\hertz} in Europe and \SIrange{902}{928 }{\mega\hertz} in North America).  Our experimentation will focus on the \SI{2.4}{\giga\hertz} frequency band. The above result in a cross-technology interference that affects the Quality-of-Service (QoS) of the network, particularly the reliability and latency that could lead to application failures. The appropriate physical layer should be considered when developing an IoT application to ensure the highest QoS. The impact is even more prominent in the case of CT-based communications~\cite{baddeley2020impact}.

\subsection{Carrier Sense Multiple Access with Collision Avoidance}\label{subsec:csma}
CSMA-CA is a random access protocol and works by sensing the channel before transmitting to avoid collisions. When the channel is occupied, it waits for a variable amount of time (back-off period) before checking again. Should a collision occur, the back-off period is increased to avoid the next packet also colliding \cite{tanenbaum_computer_networks}. The main advantage of CSMA-CA is its simplicity, as it does not require time synchronisation, scheduling between nodes, or the need to wait to join a network. However, as the number of nodes increases, it becomes saturated, and collisions become more likely. Finally, CSMA-CA operates on a predefined fixed frequency, thus is susceptible to environmental interference from nearby transmitters operating on the same or adjacent channels in the ISM band (such as WIFI).

For our experiments, we use CSMA-CA with Routing Protocol for Low-Power and Lossy Networks (RPL)~\cite{rpl}. RPL is an optimised routing protocol for wireless networks susceptible to packet loss. The routing tables are maintained as a Destination-Oriented Directed Acyclic Graph (DODAG). Each node is assigned a rank that increases as we move away from the root node. Two modes are supported, i.e. storing and non-storing modes. For the storing mode, routing tables are stored on each node, imposing a significant memory footprint in large networks and being hard to maintain consistency. For the non-storing mode, IPv6 source routing is employed. This means that routing tables are not stored in the nodes but are embedded in the source routing header. In larger networks with many hops, this can lead to increased header size.

While using the IEEE 802.15.4 channel, there is a limit to how much data can be sent. Researchers in \cite{ContikiCSMA}, introducing various modifications in the Contiki stack, reliably achieving up to \SI{45}{\Kbps} with a CSMA-CA channel, using only ten devices. However, as described, increasing the load to \SI{100}{\Kbps} introduces large latencies and significant packet loss. For our implementation, a standard CSMA-CA stack is considered, thus the throughput is expected to be less than that.

\vspace{-1mm}

\subsection{Concurrent Transmissions and Synchronous Flooding}\label{subsec:Atomic}
CT is the concept that nodes synchronously transmit in-contention with their neighbours. Synchronous Flooding (SF) builds on the idea of CT, and it can support one-to-all communication within a single flood, minimise the latency and enhance reliability. Its time-synchronised nature can help decouple network synchronisation from other network processes. An example of SF-based protocol is Atomic~\cite{baddeley2019Atomic}, which showed great results in terms of reduced latency, better reliability and energy efficiency on small testbeds. Finally, Atomic supports Multicast Protocol for Low-Power and Lossy Networks (MPL) and will be the SF protocol of use.

\begin{figure}[t]
    \centering
    \includegraphics[width=0.95\columnwidth]{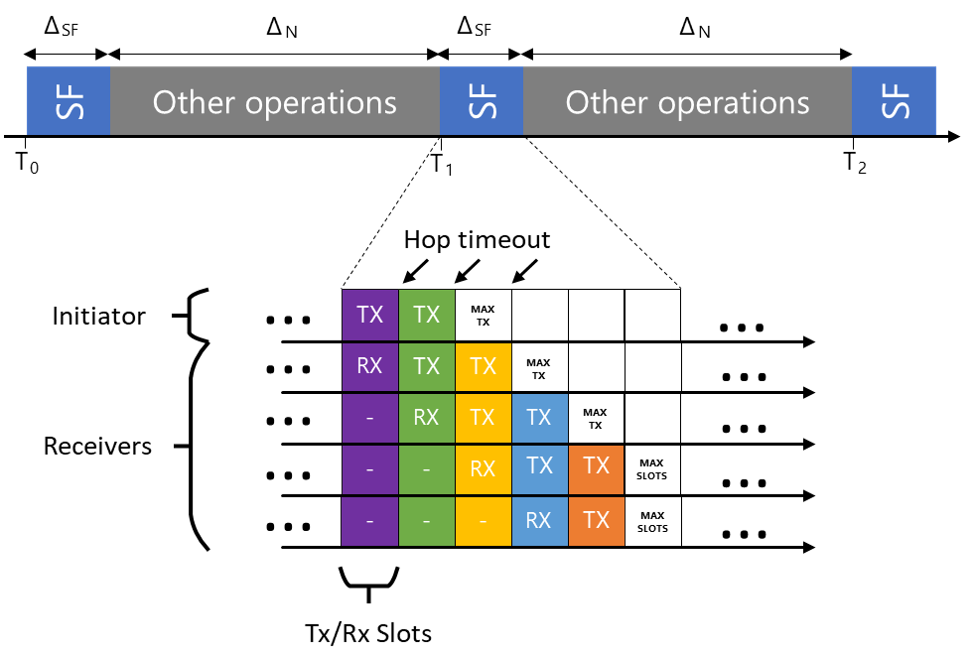}
    \vspace{-5mm}
    \caption{Example of SF used in AtomicSDN. Back-to-back transmissions flood the network with minimal latency. The MAX\_TX and MAX\_SLOTS values are a trade-off between latency, and greater temporal/frequency diversity.}
    \label{fig:Atomicsf}
\end{figure}

During an Atomic period $\Delta_{SF}$, a packet is flooded across the whole network. Each $\Delta_{SF}$ is partitioned into slots. The maximum number of slots (MAX\_SLOTS), and the maximum number of transmissions (MAX\_TX), are configured at the start of each flooding. A node transmits a packet at the start of the Atomic period to all nearby listening devices. Then, all devices that successfully received the packet re-transmit it alongside the source node. This repeats until each node has resent the packet MAX\_TX times. With each re-transmission, the data spread to exponentially more nodes. An example of the above can be seen in Fig.~\ref{fig:Atomicsf}. Starting from the initiator, a node transmits a packet repeatedly until slot MAX\_TX. Then, all others receive the packet and relay that during the next slot to all the forwarding nodes. More information about Atomic can be found in~\cite{baddeley2019Atomic}.

With such an approach, nodes receive the same symbol from multiple transmitters. This results in greater reliability for the whole network. Facilitating the propagation of control messages across the network in a flood and within dedicated control timeslots, Atomic allows an operation without the knowledge of the topology and benefits from the spatial and temporal diversity inherent within flooding protocols. Finally, Atomic uses channel hopping to increase performance, which avoids sources of interference on a fixed frequency and stops the Atomic flood from saturating one channel for too long.

\vspace{-1mm}

\subsection{UMBRELLA Testbed and Network Setup}\label{subsec:umbrellaTestbed}
As mentioned, UMBRELLA testbed~\cite{UMBRELLA} was used for our evaluation. UMBRELLA supports various IoT-related applications and use-cases. It allows the users to develop different applications and deploy them as containers on the provided edge nodes (UMBRELLA nodes). The applications supported range from air quality monitoring, street light maintenance, swarm robotics, private 5G for warehousing, logistics, and large-scale \textit{over-the-air} wireless experimentation. The large-scale wireless testbed was the particular functionality that we utilised for our performance investigation.

\begin{figure}[t]
    \centering
    \includegraphics[width=0.98\columnwidth]{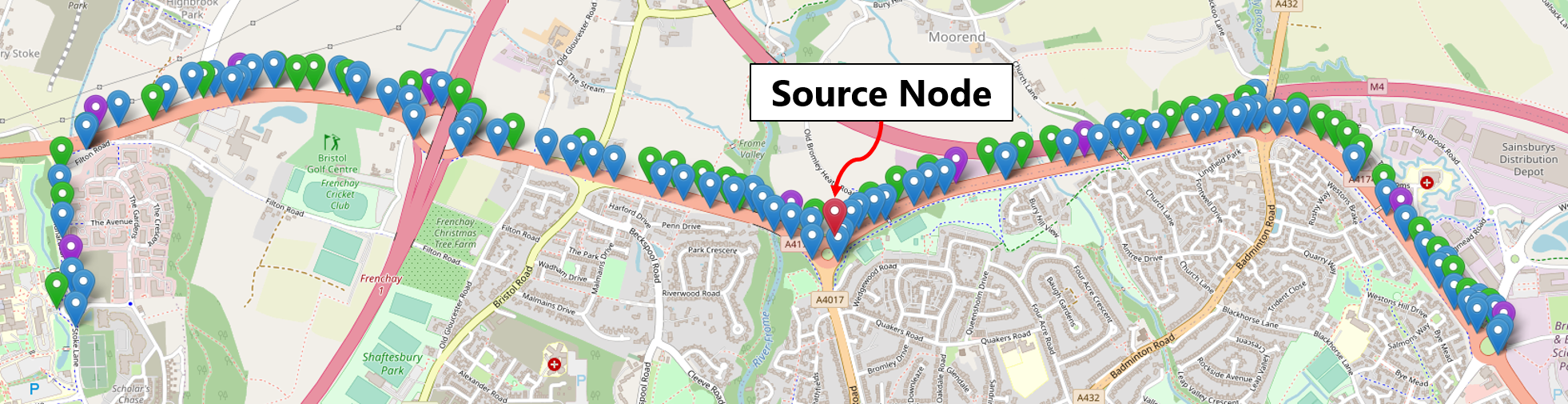}
    \vspace{-3mm}
    \caption{The UMBRELLA network. All nodes are installed on public lampposts across a road of \textapprox7.2km. The colors represent the nodes connectivity, i.e. green is fibre connected, blue is WiFi connected, and purple is fibre connected and can act as a LoRa gateway too. The red node is our experiment source node.}
    \label{fig:umbrella_network}
\end{figure}

UMBRELLA consists of \textapprox200 nodes, installed on public lampposts and buildings, spread across South Gloucestershire region in the UK. The core UMBRELLA testbed is installed across a \textapprox7.2km stretch of road (Fig.~\ref{fig:umbrella_network}).
Each node annotated in Fig.~\ref{fig:umbrella_network} is equipped with ten sensors (e.g., Bosch BME680, accelerometers, microphones, etc.) and seven network interfaces. The two interfaces used for experimentation are a Nordic Semiconductor nRF52840~\cite{nRF52840} and a Texas Instruments CC1310~\cite{CC1310}. Our investigation was based on nRF52840. Two more interfaces (WiFi and fibre one) are used for backbone connectivity. Finally, all applications are supported by a unified backend implementation. The developed platform provides the required messaging interfaces and protocols and some high-level APIs that an end-user can leverage to send requests, collect log files, or process the data and visualise them.

\begin{figure}[t]
    \centering
    \includegraphics[width=0.79\columnwidth]{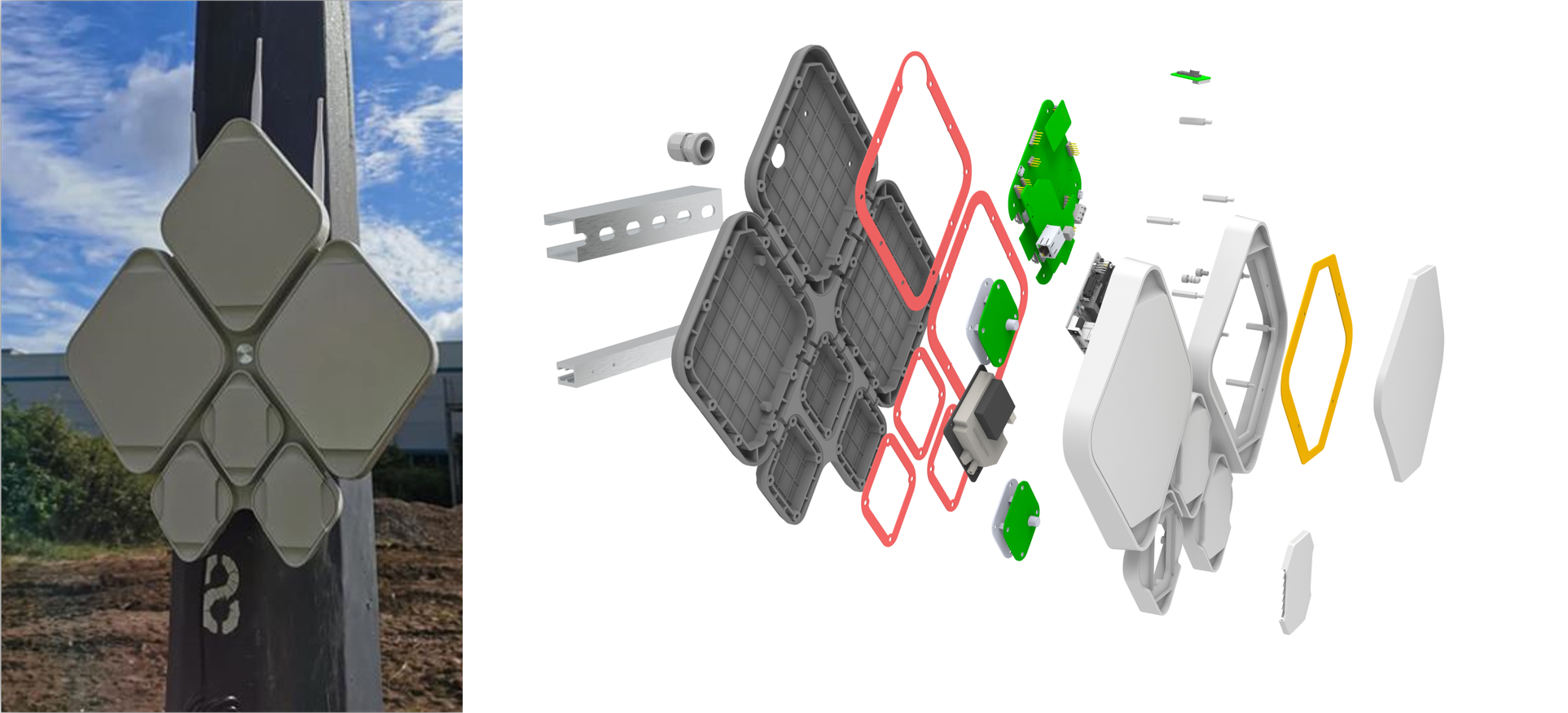}
    \vspace{-2mm}
    \caption{Umbrella Node on a lamppost, with its exploded view.}
    \label{fig:umbrella_node}
\end{figure}

Between the nRF52840 interface and its dipole antenna exists a Skyworks RF Front-End Module~\cite{skyworksFEM}, integrating a Low Noise Amplifier (LNA) and Power Amplifier (PA). This results in \SI{22}{\dB} of TX power gain, and increases RX sensitivity up to \SI{6}{\dB}, approximately doubling the range of a typical IoT device~\cite{skyworksFEM}. Fig.~\ref{fig:umbrella_node} shows an Umbrella node attached to a lamppost. Each rhombus segment contains custom PCBs, all connected to a main processing unit (Raspberry Pi 3b+ Compute Module~\cite{RPI_3_COMPUTE}).
The UMBRELLA Raspberry Pi runs Raspbian GNU/Linux 10 (buster) and a custom kernel based on ver. 4.19.95-v7+. Finally, the three dipole antennas seen in Fig.~\ref{fig:umbrella_node} are for the \SI{2.4}{\giga\hertz} nRF52840 (left), the sub-\SI{}{\giga\hertz} CC1310 (middle), and the \SI{2.4}{\giga\hertz} WiFi (right) interfaces.

\vspace{-1mm}

\section{Large Scale Experiments}\label{sec:trials}

As shown in Fig.~\ref{fig:umbrellaarchitecture}, all firmware binaries are exchanged via the wireless mesh network (the nRF52840 interface in our case). Furthermore, all nodes are also accessible via a backbone link (WiFi or fibre), used for data collection, monitoring, and experiment execution purposes.
The dedicated backbone link enhances the: 1) results collection without interference or disruption of on-going experiments, 2) available bandwidth to support large scale multi-user parallel experiments. At the beginning of each experiment, all nodes are assigned a unique ID and a role, i.e., source or receiver. The source node is the roll-out initiator (as described in Sec.~\ref{sec:system}). Only a single node is assigned this role. The receivers act as the firmware consumers. With regards to Atomic, the source node also manages the SF periods and sets up the routing and timing for all nodes. The chosen source node (Fig.~\ref{fig:umbrella_network}) is centrally located, giving us roughly an equal amount of nodes on each side of the network.

\begin{figure}[t]
    \centering
    \includegraphics[width=0.95\columnwidth]{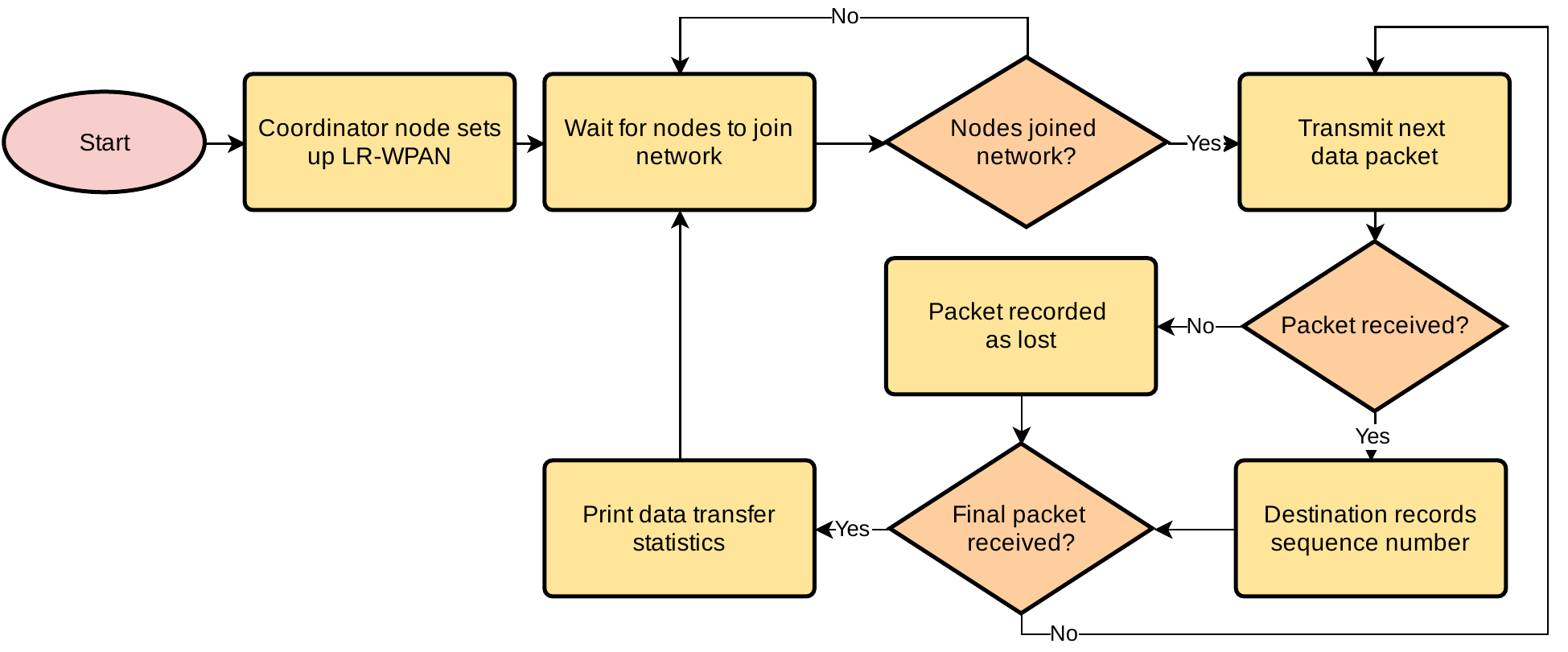}
    \vspace{-3mm}
    \caption{Experiment flowchart. The source node initiates the network setup and is responsible for the transmission of the firmware later. When all packets have been sent, the statistics of each node are collected.}
    \label{fig:flowchart}
\end{figure}

The \SI{100}{\kilo\byte} of data (emulated firmware) is discretised into packets, utilising the maximum payload size available without causing fragmentation (Tab.~\ref{table:smallScaleThoughputTable}). The remaining header bytes are used for control information, source and destination addresses and a checksum for the frame verification at the reception. Once the network connection is established, all packets are sent sequentially via the nRF52840 interface of the source node. Each packet contains a unique identifier, which is their sequence ID, so the firmware consumers are aware of which packets were received or lost. Fig.~\ref{fig:flowchart} shows the high-level structure of the procedure followed.

\vspace{-1mm}

\subsection{Transmitting and Receiving Sides}
When using Atomic, we schedule a stream of packets for transmission at the end of each Atomic period $\Delta_{SF}$ (Fig~\ref{fig:Atomicsf}). Our protocol stack is based on Contiki-NG~\cite{contiki_ng}. Thus, we use various control processes and callbacks from the existing stack. A control process scheduled on a \SI{1}{\second} timer checks the readiness of the network and the completion of a transfer. Once the connection is established, the \texttt{send\_next\_packet} function from Contiki-NG is polled, and the ``next packet'' is added to the queue and sent once the current process finishes. On the other hand, when CSMA-CA is used, we schedule a \texttt{send\_next\_packet} process in regular intervals. After preliminary experimentation, we identified the timing required for our network to maximise the throughput without causing packets to be dropped due to timeouts.

On the receiving end, a node, upon joining the network, waits for a stream of packets to arrive. A stopwatch timer is triggered when the first packet is received, and the successful receptions are monitored. We maintain two lists monitoring their unique IDs, i.e. ``packets received'' and ``packets lost''. ``Packets lost'' are the ones that arrived out of order (e.g., their sequence ID preceding the last packet received), have a checksum fail, or are not received at all. To emulate a real-world-like scenario, simplicity, and fair comparison at a large-scale, retransmissions were set to $0$. In a real-world application, this could be handled by a higher layer, e.g., passing the list of the lost packets to the transmitter and requesting the retransmission of the missing payloads. When the last packet is received, the node stops its local timer and prints its statistics.

\vspace{-1mm}

\subsection{Experimental Pipeline}
For our evaluation, an automated deployment pipeline was implemented. To simplify the experiment execution, the same firmware code is uploaded on all nodes, containing both the roll-out initiator and firmware consumer codes. Using the Contiki-NG deployment service, the role of each node is decided during run-time, i.e. hardcoding a single node (the same) as the Atomic SF initiator and the firmware roll-out source. The compiled binaries are uploaded on all nodes before flashing. Upon successful uploading, all nodes are flashed in parallel, and the experiment is initiated.

At the end of the experiment, all the results are recorded on the Raspberry Pi via the serial interface. They are later collected centrally on our UMBRELLA servers for post-processing. It was observed that the interaction with the serial interface causes a considerable delay for the SoC. Thus, it was decided to collect all the data at the end of each experiment. Furthermore, old-executed binaries can compromise the results when running sequential experiments (e.g., due to backbone network downtime nodes could be left earsplitting old data on the channel). Therefore, all nodes were flashed with a dummy binary at the end of each experiment to mitigate that. Finally, all the above processes and steps are automated with Ansible and a series of scripts to ensure smooth execution.


\vspace{-1mm}

\section{Performance Investigation}\label{sec:performance}

For our performance investigation, we started with a small-scale experiment to identify the maximum achievable datarate. For that, we ran a small-factor experiment within a controlled lab environment using both Atomic and CSMA-CA and four nodes. Particularly for CSMA-CA, in order to find the optimal transmission time, we ran multiple experiments decreasing the inter-packet time until the receiver became unreachable. The last stable configuration was used as our time interval.

For our large-scale experiment, roughly \textapprox150 nodes were used. These are the nodes seen in Fig.~\ref{fig:umbrella_network} and is a subset of the core UMBRELLA network. The rest of the nodes were excluded either due to being out-of-range or being installed inside buildings (thus not capable of forming a mesh with the outdoor nodes). All the nodes eastwards from the source are installed in a denser setup, with an average distance separation of \textapprox\SI{87}{\meter}. The nodes westwards are separated by \textapprox\SI{94}{\meter} on average. Finally, channel no. $26$ from the IEEE 802.15.4 frequency band was used for CSMA-CA, and both small- and large-scale experiments (carrier frequency of \SI{2.480}{\giga\hertz}). This channel was chosen due to its minimal observed interference after a channel sounding on the available spectrum.

\vspace{-1mm}

\subsection{Small-scale Comparison}

Tab.~\ref{table:smallScaleThoughputTable} shows a summary of the results and the experimental configuration used. The time period (measured in \SI{}{\milli\second}) is the interarrival time between two packets and is the lowest value perceived with the network still reliably transferring data. These results indicate a maximum achievable data rate under ideal-like conditions. As seen, Atomic managed to outperform CSMA-CA in terms of the throughput perceived. However, as we will see in the next sections, more conservative configuration values should be used under a real-world scenario, and thus the achieved data rate decreases.

\begin{table*}[t]
\renewcommand{\arraystretch}{1.03}
    \centering
    \caption{Small scale throughput experiment results}
    \vspace{-2mm}
    \begin{tabular}{|c|c|c|c|c|c|c|c|c|}
        \hline
        \textbf{MAC} & \textbf{PHY}  & \textbf{MAX\_TX} & \textbf{MAX\_SLOTS} & \textbf{Time Period} & \textbf{MTU} & \textbf{Payload Size} & \textbf{Throughput} & \textbf{Datarate (\%)} \\
        \hline \hline
        Atomic & BLE \SI{2}{\Mbps} & 3 & 7 & \SI{16}{\milli\second} & \SI{256}{\byte} & \SI{230}{\byte} & \SI{115.19}{\Kbps} & $5.76\%$ \\
        Atomic & BLE \SI{1}{\Mbps} & 3 & 7 & \SI{20}{\milli\second} & \SI{256}{\byte} & \SI{230}{\byte} & \SI{92.17}{\Kbps} & $9.22\%$ \\
        Atomic & BLE \SI{500}{\Kbps} & 3 & 7 & \SI{29}{\milli\second} & \SI{256}{\byte} & \SI{230}{\byte} & \SI{63.56}{\Kbps} & $12.71\%$ \\
        Atomic & BLE \SI{125}{\Kbps} & 3 & 7 & \SI{77}{\milli\second} & \SI{256}{\byte} & \SI{230}{\byte} & \SI{23.94}{\Kbps} & $19.15\%$ \\
       Atomic & IEEE 802.15.4 \SI{250}{\Kbps} & 3 & 7 & \SI{29}{\milli\second} & \SI{127}{\byte} & \SI{121}{\byte} & \SI{33.40}{\Kbps} & $13.36\%$ \\
        CSMA & IEEE 802.15.4 \SI{250}{\Kbps} & 1 & - & \SI{31}{\milli\second}  & \SI{127}{\byte} & \SI{86}{\byte} & \SI{21.98}{\Kbps} & $8.79\%$ \\ \hline
    \end{tabular}
    \label{table:smallScaleThoughputTable}
\end{table*}

\begin{figure*}[t]
    \centering
    \includegraphics[width=0.97\textwidth]{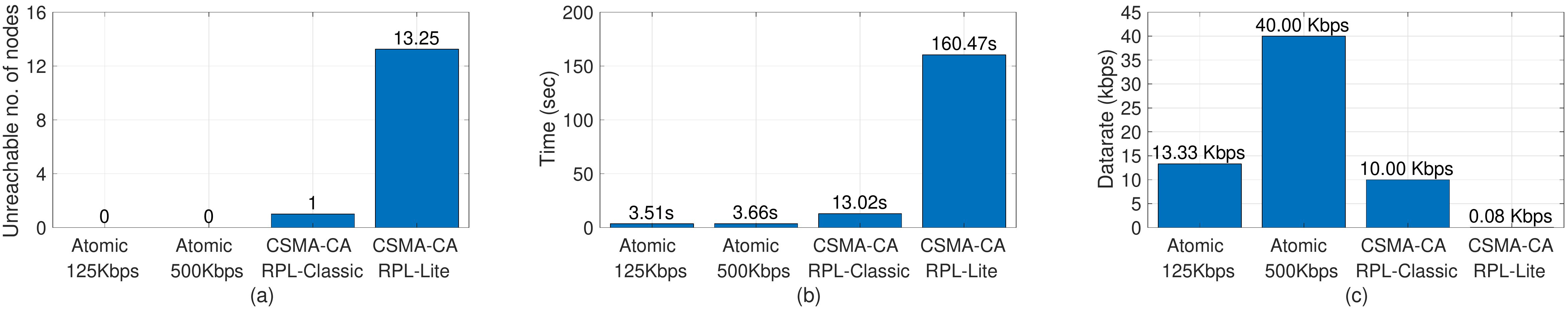}
    \vspace{-4mm}
    \caption{A comparison between Atomic and CSMA-CA. (a) shows the number of nodes non-reachable on average per experiment, (b) is the time required to form the mesh network, and (c) is the maximum achievable datarate.}
    \label{fig:allcsma}
\end{figure*}

\subsection{Atomic and CSMA-CA: a large-scale experiment}\label{subsec:atomicCSMA}

For the large-scale experiment, and using the indicative values from Tab.~\ref{table:smallScaleThoughputTable}, we adapted several parameters to fit the larger scale of the experiment. More specifically, the payload size of the CSMA-CA was reduced to \SI{50}{\Kbps}. This is because the routing table included inside the MAC header was larger, thus, more space was required to accommodate that. Also, both RPL-Classic and RPL-Lite were used for CSMA-CA. A worst-case unicast CSMA-CA scenario was chosen to approximate the effects of a MPL approach, as in~\cite{mlpCSMA}.  With what regards Atomic, the MAX\_SLOTS was increased to 8, and the MAX\_TX was increased to 12 to accommodate the increased number of nodes that take part in the experiment. These values were chosen after an exploratory investigation on the testbed. We chose the best performing pair for the given setup. Different time periods were tested, these being between {\SIrange{50}{500}{\milli\second}} with an interval of \SI{25}{\milli\second} and between {\SIrange{500}{800}{\milli\second}} with an interval of \SI{100}{\milli\second}, and two PHY were used, i.e., \SI{125}{\Kbps} and \SI{500}{\Kbps}. Finally, the denser side of the network was utilised for this experiment (all nodes eastwards from the source node).

We compare Atomic and CSMA-CA (Fig.~\ref{fig:allcsma}) against the number of nodes able to form a mesh network, the time required to join, and the achievable datarate while sending \SI{100}{\kilo\byte} of data. The nodes reached and the mesh join time are averaged across all values, while the datarate is the maximum achieved. As shown, in terms of the number of unreachable nodes, Atomic outperforms CSMA-CA (Fig.~\ref{fig:allcsma}-(a)), constantly forming a mesh network with all the available nodes. Particularly when using RPL-Lite, we observed that about $1/10\textrm{th}$ of the nodes was not part of the end network. When rolling out a firmware update, it is necessary to ensure that all the available nodes receive it. Atomic was proven a lot more reliable in such a scenario. Similar results can be seen in Fig.~\ref{fig:allcsma}-(b) where Atomic requires less time to form the network, thus achieving less downtime and being less susceptible to network changes. This reflects on the increased perceived datarate (Fig.~\ref{fig:allcsma}-(c)). As shown, Atomic, even with a lower PHY capacity (\SI{125}{\Kbps} against \SI{250}{\Kbps} for CSMA-CA), still manages to achieve better performance. For the case of PHY \SI{500}{\Kbps} the results are even more significant, showing four times higher datarate than CSMA-CA. These results show that the firmware binaries can reach the end destination faster, thus reducing the network's ``downtime'' and ``maintenance'' phases while making the roll-out less prone to errors or external factors.

\subsection{Atomic: A more thorough experimentation}
Above, we presented the dominance of Atomic compared to CSMA-CA. We now further investigate Atomic performance using the entire UMBRELLA network. A single experiment was executed throughout the entire network, but as we describe, we considered the performance perceived on the dense and sparse sides independently. As before, we used \SI{125}{\Kbps} and \SI{500}{\Kbps} PHYs. Due to the limited size of the paper, some of the results will only be described in text.

\begin{figure}[t]
    \centering
    \includegraphics[width=0.97\columnwidth]{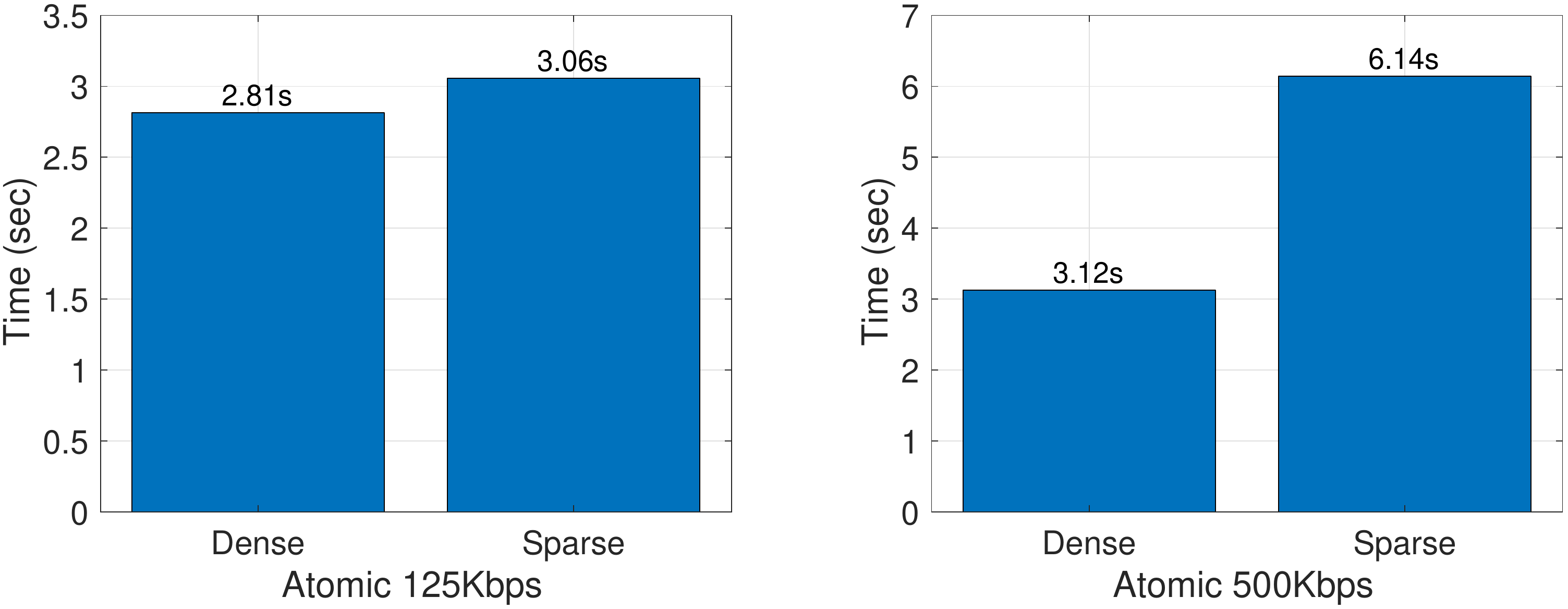}
    \vspace{-4mm}
    \caption{The average mesh join time for two Atomic PHYs and the two sides of the UMBRELLA network.}
    \label{fig:densesparce}
\end{figure}

In Fig.~\ref{fig:densesparce} we compare Atomic using two different densities and PHYs and the same periods mentioned above. The MAX\_SLOTS was set to 8, and the MAX\_TX was set to 16. For \SI{125}{\Kbps}, the mesh join time was slightly reduced compared to the experiment in Sec.~\ref{subsec:atomicCSMA}. The larger number of nodes around the source node helps with the mesh initialisation, insofar, sightly decreasing the join time. Considering the \SI{500}{\Kbps} PHY, as before, for the dense network side, the increased number of nodes around the source decreases the time required. However, when considering the sparser side, it is shown that the time is almost doubled. The increased distance separation and the higher PHY result in more hops required for each node to be reached, thus the difference in the results. The number of unreachable nodes is again zero for both PHYs, implying that all nodes were part of the mesh formation. The above findings give us a good indication of the ability of Atomic and SF to form networks under various scenarios and conditions.

\begin{figure}[t]
    \centering
    \includegraphics[width=0.97\columnwidth]{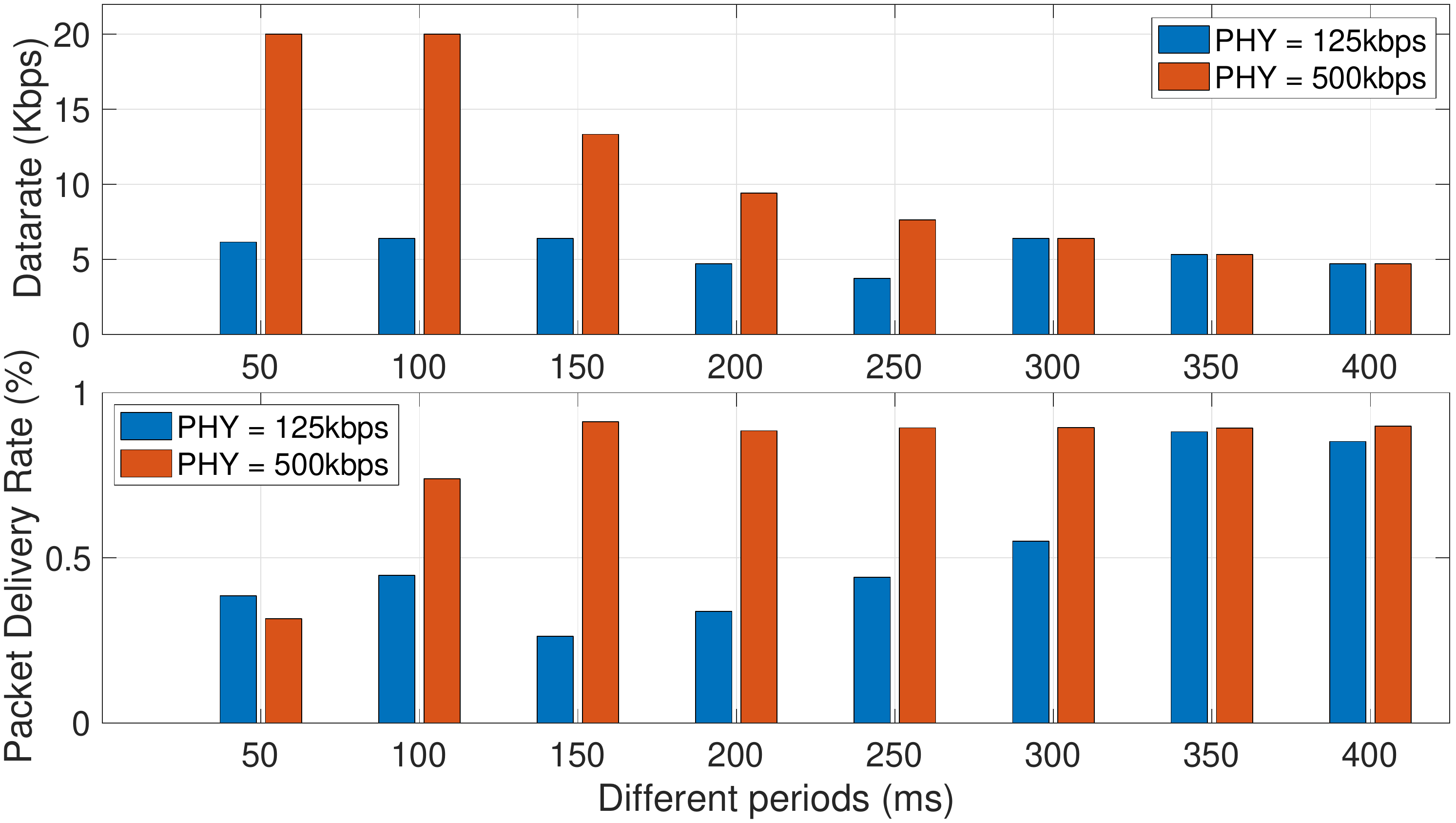}
    \vspace{-4mm}
    \caption{The maximum datarate observed and the average packet delivery rate between all the UMBRELLA nodes.}
    \label{fig:pdrthroughput}
\end{figure}

We later investigate the reliability of Atomic when rolling out the firmware update on all the consumers. As seen in Fig.~\ref{fig:pdrthroughput}, lower periods can increase the maximum datarate, however significantly reduce the reliability of the protocol. This is because short Atomic periods reduce the number of nodes participating in the Atomic SF operation. Thus, many nodes become unreachable until the next Atomic period. When the period is increased, the datarate is reduced. Allowing more time for Atomic to configure reduces the available channel capacity, increasing the time mandated for a firmware roll-out. These results show us the importance of choosing the appropriate configuration based on the available setup. By doing so, we can maximise the throughput without compromising the system stability. Focusing on the two sides of the network now (dense-sparse), it was observed that on average \textapprox6 nodes on the sparse side, and \textapprox2 on the dense one, were not able to finish the binary reception and dropped from the network during the roll-out. This was due to the operation of Atomic and the number of slots chosen. After further investigation, these nodes were identified to have very low RSSI, being at the edge of their neighbours' coverage area. Such nodes are susceptible to the way Atomic works, especially when the number of MAX\_SLOTS increases, as they cannot reliably finalise the Atomic initialisation after every Atomic period, and thus are prune to misconfigurations. The performance in such networks could be improved further by combining SF with traditional routing like in~\cite{baddeley2021tisch}.

\vspace{-1mm}

\section{Conclusions}\label{sec:conclusion}
This paper described a large-scale IoT wireless investigation using two protocols, i.e., Atomic and CSMA-CA. We adopted a typical firmware roll-out as our experimental use-case using different configurations. We used the publicly available UMBRELLA testbed for our large-scale investigation. Our results showed that Atomic and SF-like protocols outperform traditional protocols like CSMA-CA in such use-cases and larger networks. We also discovered how denser or sparser setups affect SF and Atomic and the importance of proper configuration based on the given network setup. In the future, we plan to investigate the feasibility of other protocols and extend Atomic's capabilities to accommodate considerable distance variations between neighbouring nodes.

\vspace{-1mm}

\section*{Acknowledgment}
This work is funded in part by Toshiba Europe Ltd. UMBRELLA project is funded in conjunction with South Gloucestershire Council by the West of England Local Enterprise Partnership through the Local Growth Fund, administered by the West of England Combined Authority.

\vspace{-1mm}

\bibliographystyle{IEEEtran}
\bibliography{main,IEEEabrv}

\end{document}